\documentclass[jkps,showpacs,showkeys,preprint]{revtex4}
\usepackage[dvips]{graphicx}
\usepackage{latexsym,amsmath}

\begin{document}

\preprint{\vbox{\hbox{\tt KIAS-P07026}}}

\title{Entropy of Schwarzschild Black Holes on DGP Brane}

\author{\surname{\sc Ee} Chang-Young}
  \email{cylee@sejong.ac.kr}
  \affiliation{Department of Physics, Sejong University, Seoul
    143-747, Korea}
 \affiliation{School of Physics, Korea Institute for Advanced Study,
    Seoul 130-722, Korea}
\author{\sc Daeho \surname{Lee}}
  \email{dhlee@sju.ac.kr}
  \affiliation{Department of Physics, Sejong University, Seoul
    143-747, Korea}
\author{\sc Myungseok \surname{Yoon}}
  \email{younms@sogang.ac.kr}
%  \affiliation{Center for Quantum Spacetime, Sogang
%    University, Seoul 121-742, Korea}
  \affiliation{Department of Physics, Sejong University, Seoul,
     143-747, Korea}
\date{\today}

\begin{abstract}
We study the entropy of Schwarzschild black holes on DGP brane.
The radius of event horizon on DGP brane is obtained by a numerical method.
It is smaller than that of
Einstein gravity in the conventional branch, and is larger in the
accelerated branch. However, the difference is very small. The entropy
of the black hole is calculated by using the improved brick-wall
method.
\end{abstract}


\maketitle

After Bekenstein has suggested that the entropy of a black hole is
proportional to the surface area at the event horizon
\cite{bekenstein}, there have been several studies for the statistical
origin of the entropy\cite{hawking,zt,fn}. This can be understood by
using a brick-wall method\cite{thooft} or an \textit{improved} brick-wall
method\cite{hzk,kpsy}. In the improved brick-wall method, a thin
layer at the horizon is used instead of an infinite spherical box.  As
for a black hole on the Dvali-Gabadadze-Porrati(DGP) model\cite{dgp},
an exact expression of metric has not been obtained. In this paper, we
first get the radius of event horizon numerically, then, based on this
data about horizon, calculate the entropy of the black hole by using an
improved brick-wall method.

The action in the DGP model is given by\cite{dgp}
\begin{equation}
  \label{action}
  S = M_*^3 \int d^5 x \sqrt{-g} R + M_P^2 \int d^4 x \sqrt{-\tilde{g}}
  \tilde{R}
\end{equation}
where the tilde denotes the quantities on the four-dimensional
world-volume of the brane. A cross-over scale is defined by $r_c =
m_c^{-1} = M_P^2/(2M_*^3)$. We consider a spherically symmetric and static
black hole on the brane. Then, the five-dimensional line element can be
written as
\begin{equation}
  \label{metric}
  ds_{(5)}^2 = -e^{-\lambda}dt^2 + e^{\lambda} dr^2 + r^2 d \Omega^2
  +\gamma dr dy + e^\sigma dy^2,
\end{equation}
where $\lambda$, $\gamma$, and $\sigma$ are functions only for $r$ and
$y$ and the brane is located at $y=0$. The equations of motion from
the action (\ref{action}) with Israel's junction conditions reduce to
the equation for $\lambda$ on the brane as follows\cite{gi:prd}:
\begin{equation}
  \label{diffU}
  U^2_z-4(1+U)U_z -8U(2+U)= 0,
\end{equation}
where $z=\ln(r/r_0)$, $U = -2 \exp(-2z)P_r / 3m^2_cr^2_0$, $P_r =
dP/dr$, and $P(r)=r(1-\exp(-\lambda))|_{y=0}$ ($r_0$ being an
arbitrary constant). Two solutions of
Eq.\ (\ref{diffU}), \textit{regular} (or \textit{conventional}) and
\textit{accelerated} branches, are given by\cite{gi:prd}
\begin{eqnarray}
  k_1r \!\! &=&\!\! R(U)
     = \left[ -\frac{(1 + 3U + f)}{U^2 (3 + 3U +
       \sqrt{3}f)^{2\sqrt{3}}(- 5 - 3U + f)} \right]^{\frac18},
       \label{ktoRegulU} \\
  k_2r \!\!&=&\!\! A(U)
     = \left[ - \frac{(- 5 - 3U + f)(-3 - 3U -
       \sqrt{3}f)^{2\sqrt{3}}}{(U + 2)^2( 1 + 3U + f)}
       \right]^{\frac18}, \label{ktoAccelU}
\end{eqnarray}
where $k_1$ and $k_2$ are constants of integration and $f = \sqrt{1 +
  6U + 3U^2}$. By applying boundary conditions by $P(0)=r_M$ and
$P(\infty)=0$ in the regular branch and $P(0)=r_M$ and $P(r)\approx
m^2_cr^3$ for large $r$ in the accelerated branch, $k_1$ and $k_2$ are
determined as $2(r_*k_1)^3=c_1 \approx 0.43$ and $2(r_*k_2)^3 = c_2
\approx 4.41$, where $r_*=(r_Mr^2_c)^{1/3}$\cite{gi:plb}.
From $P(r)=-\frac{3}{2} m^2_c \int dr\ r^2 U(r)$, we obtain
\begin{eqnarray}
  P(r)\!\! &=& \!\! r_M\left[1 - \frac{1}{c_1}U R(U)^3 - \frac{1}{c_1}
    \int_{U}^{\infty} \! dU R(U)^3 \right], \label{RegulP} \\
  P(r) \!\!&=&\!\! r_M\left[1- \frac{1}{c_2} UA(U)^3 + \frac{1}{c_2}
    \int_{-\infty}^{U} \! dU A(U)^3 \right], \label{AccelP}
\end{eqnarray}
which correspond to the regular and accelerated branches,
respectively.

\begin{figure}[pt]
  \includegraphics[width=0.5\textwidth]{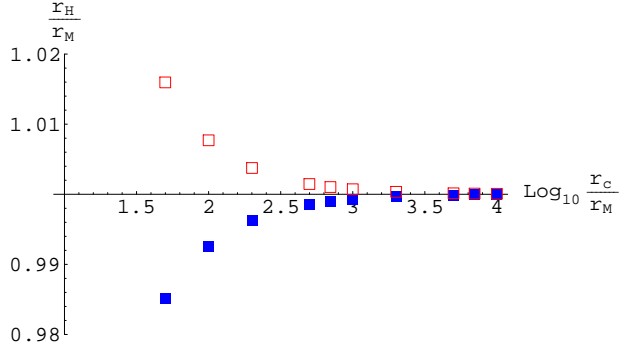}
  \caption{Filled boxes and unfilled boxes are the ratios of $r_H$ to
    $r_M$ in the regular and the accelerated branches, respectively,
    where $r_H$ is a horizon of a black hole on DGP brane and $r_M$ is
    a horizon of a Schwarzschild black hole in Einstein gravity. It
    shows that $r_H$ approaches $r_M$ when $r_c$ is much larger than
    $r_M$.}
  \label{fig:horizon}
\end{figure}

Since $ P(r_H)=r_H $ at the horizon, we can obtain $U_H \equiv
U(r_H)$ from Eqs.\ (\ref{ktoRegulU}) -- (\ref{AccelP}). Substituting
the corresponding $U_H$'s into Eqs.\ (\ref{ktoRegulU}) and
(\ref{ktoAccelU}), the horizons are calculated from $r_H/r_M  =\left[
2(r_c/r_M)^2 /c_1 \right]^{1/3}R(U_H)$ for the regular branch and
$r_H/r_M = \left[2 (r_c/r_M)^2/c_2\right]^{1/3}A(U_H)$ for the
accelerated branch. The results for the regular
and the accelerated branches are shown in Fig.~\ref{fig:horizon}.
Note that the horizon can be written as $r_H = r_M + \xi$, where $\xi$
should be very small because $r_c/r_M$ is very large. This fact will be
taken into account for calculating the entropy.

In order to find the entropy of a black hole, we would rewrite the
line element (\ref{metric}) on the brane as $ds^2 = -f dt^2 + f^{-1}
dr^2 + r^2 d\Omega^2$, where $d\Omega$ denotes a solid angle. The
asymptotic behavior of the metric is obtained as
\begin{eqnarray}
  f(r) &\approx& 1 - \frac{r_M}{r} + b\alpha m_c^2 r^2 \left(
    \frac{r_\ast}{r} \right)^{2(\sqrt{3}-1)}, \ \textrm{for $r \ll
    r_\ast$} \label{f:H} \\
  f(r) &\approx& 1 - \frac{\tilde{r}_M^2}{r^2} - b m_c^2 r^2, \qquad
    \qquad\qquad\ \textrm{for $r \gg r_\ast$} \label{f:inf}
\end{eqnarray}
where $\alpha \approx 0.84$ and $b=1$ and $b=-1$ correspond to the
conventional branch and the accelerated branch, respectively,
\cite{gi:prd,gi:plb}.

Since $a \equiv b\alpha m_c^2 r_*^{2(\sqrt{3}-1)}$ is very small,
there exists a solution $r_H$ such that $f(r_H) = 0$, where $r_H = r_M
+ \xi$ with $|\xi| \ll r_M$. The fact that $\xi$ is very small
compared to $r_M$ appears in Fig.~\ref{fig:horizon}.  From $f(r_H) =
f(r_M + \xi) = 0$, we obtain $\xi \approx -a r_M^{5-2\sqrt{3}}$. Note
that the radius of the event horizon is smaller than that of Einstein
gravity in the conventional branch, and it is larger in the
accelerated branch. Now, the Hawking temperature is given by $T =
\beta^{-1} = \kappa/(2\pi)  \approx \gamma/(4\pi r_H)$, where $\kappa$
is a surface gravity of the black hole and $\gamma \equiv 1 - \xi/r_H
+ (4-2\sqrt{3}) a r_H^{4-2\sqrt{3}} \approx 1 -
(5-2\sqrt{3})\xi/r_H$.

In order to calculate the entropy of a given system in the
brick-wall method, we consider a quantum gas of scalar particles
confined within a box near the horizon of a black hole and introduce
a cut-off parameter\cite{thooft}. The free scalar field is assumed to
satisfy the Klein-Gordon equation, $(\Box+m^2)\Psi = 0$, with
boundary conditions $\Psi(r_H + h) = \Psi(L) = 0$,
where $r_H$ is the horizon, $m$ is the mass of a scalar field, $h$ is
an infinitesimal cut-off parameter, and $r_H + h$ and $L$ represent
the inner and the outer walls of a ``spherical'' box, respectively.
Suppose that this system is in thermal equilibrium at a
temperature $T$ with an external reservoir. Using $\Psi =
\exp(-i\omega t) \psi(r,\theta,\varphi)$, the Klein-Gordon equation is
reduced to
\begin{equation}
  \label{KG:metric}
  \frac{\partial^2\psi}{\partial r^2} + \left(\frac{2}{r} +
    \frac{1}{f} \frac{\partial f}{\partial r} \right) \frac{\partial
    \psi}{\partial r} + \frac{1}{f} \left[ \frac{\omega^2}{f} - m^2 +
    \frac{1}{r^2} \left( \frac{\partial^2}{\partial\theta^2} +
      \cot\theta \frac{\partial}{\partial\theta} + \csc^2 \theta
      \frac{\partial^2}{\partial\varphi^2} \right) \right]\psi = 0.
\end{equation}
By the WKB approximation, using $\psi \sim \exp
[iS(r,\theta,\varphi)]$, we obtain $p_r^2 = f^{-1} [\omega^2/f - m^2
- p_\theta^2/r^2 - p_\varphi^2/(r^2\sin^2\theta)]$, where $p_r =
\partial S/\partial r$, $p_\theta = \partial S/\partial \theta$, and
$p_\varphi = \partial S/\partial \varphi$. Then, $p^2 = g^{rr}p_r^2 +
g^{\theta\theta} p_\theta^2 + g^{\varphi\varphi} p_\varphi^2 =
\omega^2/f - m^2$. The number of quantum states with energy
not exceeding $\omega$ can be written as
\begin{equation}
  n(\omega) = \frac{1}{(2\pi)^3} \int dr d\theta d\varphi dp_r
       dp_\theta dp_\varphi
    = \frac{2}{3\pi} \int dr \, \frac{r^2}{\sqrt{f}} \left(
      \frac{\omega^2}{f} - m^2 \right)^{\frac32}. \label{n}
\end{equation}
The free energy is given by
\begin{equation}
  F = - \int_{m\sqrt{f}}^\infty d\omega \,
        \frac{n(\omega)}{e^{\beta\omega} - 1}
    = -\frac{2}{3\pi\beta^4} \int dr \frac{r^2}{f^2} \int_{x_0}^\infty
    dx \, \frac{(x^2 - x_0^2)^{3/2}}{e^x - 1},
  \label{F}
\end{equation}
where $x=\beta\omega$ and $x_0 = \beta m \sqrt{f}$.

The degrees of freedom are dominant near horizon. Thus, we consider a thin
layer with $L=r_H + h + \delta$, where $\delta$ is small. Since $x_0$
goes to zero near horizon, we obtain
\begin{equation}
  \label{F:layer}
  F = -\frac{2}{3\pi\beta^4} \int dr\, \frac{r^2}{f^2} \int_0^\infty
  dx\, \frac{x^3}{e^x-1} \approx -\frac{2\pi^3}{45\beta^4} \int_{r_H+
    h}^{r_H + h + \delta} dr \, \frac{r^2}{f^2}.
\end{equation}
The metric near the horizon can be written as $f(r) \approx
\gamma (1-r_H/r)$. Therefore, the free energy can be calculated as
\begin{equation}
  \label{F:final}
  F \approx - \frac{2\pi^3 r_H^4}{45\beta^4\gamma^2}
  \frac{\delta}{h(h+\delta)},
\end{equation}
and the entropy becomes
\begin{equation}
  \label{S}
  S = \beta^2 \frac{\partial F}{\partial\beta}
    \approx \frac{8\pi^2 r_H^4}{45\beta^2\gamma^2}
    \frac{\bar{\delta}^2}{\bar{h}^2(\bar{h}^2+\bar{\delta}^2 )},
\end{equation}
where $\bar{h} \equiv \int_{r_H}^{r_H + h} \sqrt{g_{rr}}\, dr$,
$\bar{\delta} \equiv \int_{r_H + h}^{r_H + h + \delta} \sqrt{g_{rr}}\,
dr$.  Note that $\bar{h}$ and $\bar{\delta}$ are the cut-off parameter
and the thickness of the layer, respectively.  Thus, we can write the
entropy as
\begin{equation}
  \label{S:A}
  S \approx \frac14 {\cal A} \cdot \frac{1}{90\pi}
  \frac{\bar{\delta}^2}{\bar{h}^2(\bar{h}^2+\bar{\delta}^2 )},
\end{equation}
where ${\cal A} = 4\pi r_H^2$. When the cut-off parameter $\bar{h}$
is chosen as $\bar{\delta}^2 / [\bar{h}^2(\bar{h}^2 + \bar{\delta}^2)
] = 90\pi/G$, Eq.\ (\ref{S:A}) agrees with the Bekenstein-Hawking
entropy $S_{\rm BH} = {\cal A}/4G$\cite{bekenstein,hawking}. 

One may consider the cut-off introduced in the brick-wall method a bit
ad hoc. In Ref$.$\cite{dlm,kksy}, this point was criticized and the
Pauli-Villars regularization scheme was used to replace the
cut-off. In Refs$.$\cite{jp}, the entropy was evaluated based on the
notion of entanglement entropy. Below, we evaluate the entropy in the
thin layer using the Pauli-Villars scheme and compare it with our result
(\ref{S:A}). Introducing the five regulator fields ($i=1,\dots,5$)
besides the original field ($i=0$)\cite{dlm}, the total free energy
becomes 
\begin{equation}
  \label{F:regul}
  \bar F \approx - \frac{2r_H^2}{3\pi\gamma^2} \sum_{i=0}^5 \Delta_i
  \int_0^\infty \frac{d\omega}{e^{\beta\omega} - 1} \int_{h'}^{L'} ds
  \frac{(\omega^2 - \gamma m_i^2 s)^{3/2}}{s^2(1 - s)^4},
\end{equation}
where $\Delta_0 = \Delta_3 = \Delta_4 = +1$ for the commuting fields
and $\Delta_1 = \Delta_2 = \Delta_5 = -1$ for the anticommuting
fields and the integration is taken for values where the square root
is real. Setting $h'=0$ and focusing only on the divergent
contributions at the horizon, we obtain the entropy
\begin{equation}
  \label{S:regul}
  \bar S \approx \frac{\pi r_H^3}{3\beta} \left( B +
    \frac{32\pi^2}{15\beta^2\gamma^2} A \right) = \frac{\cal A}{4}
  \frac{B\gamma}{12\pi} + \frac{A\gamma}{90},
\end{equation}
with the same $A$ and $B$ coefficients given in Ref$.$\cite{dlm}. As
it was discussed in Ref$.$\cite{dlm}, the second term including $A$
corresponds to the 1-loop correction to the quadratic-curvatre terms
of the gravitational acton, and is irrelevant to our case. 
The first term corresponds to the 1-loop renormalization of the
Bekenstein-Hawking entropy  $S = {\cal A}/4G_R$, where the
renormalized Newton's constant $G_R$ is given by  $1/G_R =1/G_B +
B\gamma/12\pi$ with the bare Newton's constant $G_B$. 
Note that our result slightly differs from the result of
Ref$.$\cite{dlm} by a factor $\gamma$. 

%%%%%%%%%%%%%%%%%%%%%%%%%%%%%%%%%%%%%%%%%%%%%%%%%%%%%%%%%%%
%%%%%%%%%%%%           Acknowledgments           %%%%%%%%%%
%%%%%%%%%%%%%%%%%%%%%%%%%%%%%%%%%%%%%%%%%%%%%%%%%%%%%%%%%%%

%\noindent
%{\large
%{\bf
\section*{Acknowledgments}
 E.\ C-Y.\ and D.\ L.\ were supported by the Korea
Research Foundation Grant funded by the Korean Government(MOEHRD),
KRF-2006-312-C00498.
E.\ C-Y.\ thanks KIAS for hospitality during the time that this
work was done.
M.\ Yoon was supported by the Korea Research
Foundation Grant funded by the Korean Government
(MOEHRD) (KRF-2005-037-C00017).
 \\

%\vspace{5mm}
%%%%%%%%%%%%%%%%%%%%%%%%%%%%%%%%%%%%%%%%%%%%%%%%%%%%%%%%%%%%%
%%%%%%%%%%%%%%%             References       %%%%%%%%%%%%%%%%
%%%%%%%%%%%%%%%%%%%%%%%%%%%%%%%%%%%%%%%%%%%%%%%%%%%%%%%%%%%%%
%\noindent
%{\large \bf References} \\

\end{document}